\documentclass{PoS}

\usepackage{subfigure}
\usepackage{amsmath}

\title{Calculating $B_K$ using HYP staggered fermions}

\ShortTitle{Calculating $B_K$ using HYP staggered fermions}

\author{Taegil Bae, Hyung-Jin Kim,
Jongjeong Kim, Kwangwoo Kim, Boram Yoon, and \speaker{Weonjong Lee} \\
Frontier Physics Research Division and
Center for Theoretical Physics, \\
Department of Physics and Astronomy,
Seoul National University, Seoul, 151-747, South Korea \\
E-mail: \email{wlee@phya.snu.ac.kr}}

\author{Chulwoo Jung \\
Physics Department, Brookhaven National Laboratory,
Upton, NY11973, USA \\
E-mail: \email{chulwoo@bnl.gov}}

\author{Stephen R.~Sharpe \\
Department of Physics, University of Washington, Seattle,
WA 98195-1560, USA \\
E-mail: \email{sharpe@phys.washington.edu}}

\abstract{We give an update on our calculation of $B_K$ using HYP-smeared
valence staggered quarks. We have results for $B_K$ at tree-level
on several coarse MILC lattices
($a\approx 0.12\,$fm)
and one of the fine lattices ($a\approx 0.09\,$fm), using 10 light
valence quarks ranging down to $m_s^{\rm phys}/10$. 
We have generalized staggered chiral perturbation theory to our mixed action
setup, and outline the results.
We explain our present fitting strategy,
and give some preliminary results.}

\FullConference{The XXVI International Symposium on Lattice Field Theory\\
		 July 14-19 2008\\
		 Williamsburg, Virginia, USA}

\begin{document}

\section{Introduction}
An accurate result for the kaon $B$-parameter
is important both for its phenomenological
impact~\cite{Lellouchlat08}
and as a bellwether of the success in incorporating chiral symmetry
and controlling systematics in lattice calculations.
Calculations using several different fermion methods are underway,
with the present best result using domain-wall fermions~\cite{DWFBK}.
We are pursuing a calculation using improved staggered fermions.
This has the advantage of being computationally cheap, but the
challenge of dealing with the effects of taste-breaking in a context
where there is operator mixing~\cite{ref:sharpe:1}.

We use the standard staggered action with HYP-smeared links for our
valence fermions, which reduces taste-breaking by a factor of 3
compared to asqtad quarks~\cite{ref:wlee:11,Baelat08}.  We use the
MILC lattices generated with $2+1$ flavors of asqtad sea quarks.  For
the coarse MILC lattices, on which we focus here, the resultant
taste-breaking in the pion masses, while reduced compared to asqtad
quarks, remains large enough that we must use the standard
power-counting of staggered chiral perturbation theory, in
which $a^2 \approx p^2$.  The complications that this introduces have
been explained in Ref.~\cite{ref:sharpe:1}.

Our method for calculating $B_K$ using wall sources is explained in
Ref.~\cite{msBKpaper}. For each lattice we use 10 valence quark masses
running from $\approx m_s^{\rm phys}$ down to $\approx m_s^{\rm
  phys}/10$ in equal steps, and we calculate $B_K$ and $m_K$ for the
Goldstone taste for all 55 quark-mass combinations.  Results for
one-loop matching factors using the mixed action are not yet complete
so we use tree-level matching. For this, and other reasons to be
explained, all results obtained here should be regarded as very
preliminary. They are essentially our first pass at fitting the data,
which we are using to inform subsequent fitting and to determine
where improvements in statistics are needed.

\section{Staggered chiral perturbation theory for a mixed action}
\label{sec:sxpt}
Staggered chiral 
perturbation theory (S$\chi$PT) \cite{ref:wlee:1,ref:bernard:1}
incorporates discretization errors into the chiral expansion,
and in particular includes the effects of taste-breaking
and rooting. For our lightest kaons,
such effects are comparable to those coming from
the explicit chiral symmetry breaking due to quark masses,
and thus enter S$\chi$PT at LO. A major effect in S$\chi$PT is
that chiral loops (which begin at next-to-leading-order [NLO])
must be evaluated with the masses of the pions of the
appropriate tastes, rather than a common mass.
For $B_K$, which involves an insertion of the weak Hamiltonian,
one must also deal with mixing between operators
having different tastes. This leads to a significantly larger
number of unknown coefficients multiplying NLO terms than are
present in the continuum, as explained in Ref.~\cite{ref:sharpe:1}.

The S$\chi$PT analysis of Ref.~\cite{ref:sharpe:1} does not apply
directly to our set-up, however, because we use a mixed action.
Following the methods developed in other mixed-action
contexts~\cite{MAChPT}, we have generalized the results of
Ref.~\cite{ref:sharpe:1} to our setup. This turns out to be straightforward.
Here we only give a summary---further details will be presented in
Ref.~\cite{ref:future}.

There are three classes of effects resulting from using a mixed
action.  The first, which corresponds to the short-distance parts of
sea-quark loops, is simply that the coefficients multiplying various
terms in the S$\chi$PT expression for $B_K$ will change. Since these
coefficients were previously unknown, however, this has no practical
impact---one trades one set of unknown coefficients for another.

The second class comes from loop diagrams involving mixed
pions---those composed, say, of a valence (HYP) quark and a sea
(asqtad) antiquark. These, however, are absent for $B_K$ at NLO.

The third class involves loop diagrams with pions composed
of sea-quarks alone. For $B_K$ at NLO these loops all
include ``hairpin'' vertices,
for this is the only way in which the long-distance part of sea-quark loops
can enter. The effect of using a mixed action then boils down to the
need to distinguish between three types of
hairpin vertices---valence-valence, valence-sea and sea-sea---all
of which would be the same if the same valence and sea quarks
were being used. It turns out that this has an impact only
for tastes V and A (the taste singlet hairpin coupling to a particle
which is being integrated out anyway), so that one ends up with
6 hairpin parameters (3 for each of two tastes) instead of 2.
We name the hairpin parameters $\delta_B^{'vv}$,
$\delta_B^{'vs}$ and $\delta_B^{'ss}$, with $B=A$ or $V$,
and the superscript indicating the types of quark involved.
They appear at NLO in the combinations
\begin{equation}
\delta_B^{MA1} = (\delta_B^{'vs})^2/\delta_B^{'ss}\,, \quad
\delta_B^{MA2} = \left[\delta_B^{'vv} \delta_B^{'ss}-
(\delta_B^{'vs})^2\right]/\delta_B^{'ss}\,.
\end{equation}
These only enter in the expressions for non-degenerate quarks---those
for degenerate quarks in the kaon do not involve hairpin vertices.
Specifically, in the contribution
denoted ${\cal M}_{\rm disc}^{PQ}$ in Ref.~\cite{ref:sharpe:1},
and given in eq.~(50) of that paper, one must, for both
$B=V$ and $A$, make the substitution
\begin{equation}
\delta'_B \to \delta_B^{MA1} \,,
\end{equation}
and add the following new term 
\begin{equation}
a^2 \delta_B^{MA2} \frac{2 C_\chi^{2B}+C_\chi^{3B}}{\pi^2 f^4}
\left(2 \frac{\ell(Y_B)-\ell(X_B)}{Y_B-X_B}+
\widetilde\ell(X_B)
+\widetilde\ell(Y_B)\right)\,.
\label{eq:MA2}
\end{equation}
Here we use the notation $X_B$ ($Y_B$)
for the mass-squared of the flavor non-singlet
valence pion with taste $B$ and composition $\bar x x$ ($\bar y y$),
where the kaon itself has the composition $\bar x y$
(and taste $P$). In the notation
of Ref.~\cite{ref:sharpe:1} $X_B=m_{X_B}^2$
and $Y_B=m_{Y_B}^2$. 
The functions $\ell$ and $\widetilde\ell$ are chiral logarithms,
and are defined in Ref.~\cite{ref:sharpe:1}.

The original hairpins, $\delta_B^{'ss}$, enter through their
(unchanged) contributions to the masses of the flavor-singlet mesons,
$\eta_B$ and $\eta'_B$.

The hairpin parameters are a measure of taste-symmetry breaking,
and we expect them to satisfy a similar hierarchy to that
we observe in the pion spectrum, namely 
$\delta_B^{'vv}/\delta_B^{'ss}\approx 1/3$. In words, hairpins correspond
to quark-antiquark pairs communicating through intermediate gluons,
and the taste-breaking component is, by construction, reduced for
our HYP-smeared valence quarks. 
Based on this argument we also expect that
$(\delta_B^{'vs}/\delta_B^{'ss})^2\approx 1/3$. 
Combining these expectations we find
\begin{equation}
\delta_B^{MA1}\approx \delta_B^{'ss}/3
\quad {\rm and} \quad
\delta_B^{MA2}\approx 0\,.
\end{equation}
If these expectations are accurate, then using
a mixed action has essentially no impact on chiral
fitting, because the new terms proportional to
$\delta_B^{MA2}$ [eq.~(\ref{eq:MA2})] can be dropped, and the terms
proportional to
$\delta_B^{MA1}$ were present anyway with
an unknown coefficient.

\section{Fitting strategy}

The NLO expression for $B_K$ in partially-quenched
mixed-action S$\chi$PT takes the form
\begin{equation}
B_K = \sum_{i=1}^{16} c_i f_i\,,
\label{eq:fitform}
\end{equation}
in which $f_i$ are known functions and $c_i$ are
coefficients to be determined. Of the 16 unknown
coefficients, 1 ($c_0$) appears at LO (and is the
value of $B_K$ when $a\to 0$ and then the chiral limit
is taken), 4 are the NLO low-energy constants (LECs) present
in the continuum, and the remaining 11 are LECs due to
lattice artifacts and truncated perturbative matching.
For degenerate quarks these numbers reduce to 9 coefficients $=$
1 LO $+$ 3 continuum LECs $+$ 5 ``lattice LECs''.
This counting is for a single lattice spacing---the
dependence of the artifacts on $a$ 
involves a mix of $a^2$, $\alpha^2$ and $\alpha^2 a^2$~\cite{ref:sharpe:1}.

The functions $f_i$ depend on the masses
of the flavor-non-singlet valence pions of all tastes and
all compositions 
($\bar xx$, $\bar yy$ and $\bar xy$),
which we determine as part of our calculation~\cite{Baelat08}.
In addition, 
we need the
masses of the sea-quark $\bar\ell\ell$ and $\bar s s$ flavor non-singlet
pions with taste $V$, $A$ and $I$, and of the 
flavor singlet $\eta_B$ and $\eta'_B$ for $B=V$ and $A$.
These we take from the results of the MILC collaboration
(including the axial and vector hairpin vertices $\delta_B^{'ss}$)
\cite{ref:MILC:2004}.
The final inputs we need are $a$ (we use the MILC values) and
$f$ (which, for the moment, we simply set to $f=132\,$MeV).

Although we have 10 degenerate and 45 non-degenerate data points 
on each lattice, a direct fit to eq.~(\ref{eq:fitform}) is difficult,
since many of the fit functions are similar.
As a first stage, therefore, we use
only a few representative ``lattice'' contributions,
while keeping all the continuum terms. 
We also use Bayesian priors to constrain some terms, since we
know their orders of magnitude.

The fit functions we use are, firstly,
\begin{equation}
f_1  = 1 + \frac{3}{8 f^2 G} [{\cal M}_{conn} + 
{\cal M}_{disc}]^{\rm continuum}\,,
\end{equation}
where $G=m_{xy,P}^2$,
${\cal M}_{conn, disc}$ come from quark connected/disconnected
1-loop diagrams and are given in Ref.~\cite{ref:sharpe:1},
and the superscript indicates that we keep only the
contributions from chiral operators present in the
continuum. We do, however, include taste breaking in the pion masses
in these contributions. We set the scale in the chiral logarithm
to $\mu=1\,$GeV. 
Next we include the continuum analytic terms
(with the $\chi$PT scale set to $\Lambda=1\,$GeV):
\begin{align*}
f_2 & = G / \Lambda^2\,,
& f_3 & = ( G / \Lambda^2 )^2\,,
& f_5 & = (X_P - Y_P)^2/ (G \Lambda^2)\,,
\end{align*}
Note that $f_3$ is a NNLO contribution, but is needed
to fit our data up to the highest quark masses.

Finally, we include three representative discretization terms.
We use the contribution to ${\cal M}_{conn}$ containing
taste-$T$ pions, which contains two parts with independent
coefficients~\cite{ref:sharpe:1}:
\begin{eqnarray}
f_6  &=& \frac{3}{8 f G }
\left(\ell(X_T)+\ell(Y_T)-2\ell(m_{xy,T}^2)\right)
\end{eqnarray}
and the remainder which we call $f_4$.
We also include the likely dominant contribution to
${\cal M}_{disc}$, which is that proportional to $\delta_A^{MA1}$.
We set $c_7=(2 C_\chi^{2A}+C_\chi^{3A}) a^2 \delta_A^{MA1}/(\pi^2 f^4)$,
and $f_7$ is the coefficient of this term in eq.~(50)
of Ref.~\cite{ref:sharpe:1}.
Note that $f_{5,6,7}$ contribute only for non-degenerate quarks.

Based on the power-counting of Ref.~\cite{ref:sharpe:1},
the coefficients should have the magnitudes:
\begin{align}
c_1 & \approx c_2 \approx c_3 \approx c_5 \sim {\cal O}(1)
\nonumber \\
c_4 & \approx c_6 \approx \Lambda_{QCD}^2 (a \Lambda_{QCD})^2 = 0.003 
\quad \mbox{to} \quad \Lambda_{QCD}^2 \alpha_s^2 
= 0.01 \ 
\mbox{GeV${}^2$ on the coarse lattices.}
\label{eq:guess:1}\\
c_7 & \approx \Lambda_{QCD}^4 (a \Lambda_{QCD})^2 = 0.0003
\quad \mbox{to} \quad \Lambda_{QCD}^4 \alpha_s^2
= 0.001 \ 
\mbox{GeV${}^4$ on the coarse lattices.}
\nonumber 
\end{align}
Here, we use $\Lambda_{QCD} \approx 0.3$ GeV and
$\alpha_s = \alpha_{\overline{\rm MS}}(\mu=1.6 GeV) \approx 0.36$.
The smallness of $c_{4,6,7}$ is somewhat offset by the fact
that $f_{4,6,7}$ are logarithmically divergent in the chiral limit.

%

\section{Examples of fits}
\label{sec:data}
\begin{figure}[t!]
\centering
    \includegraphics[width=.45\textwidth]
                    {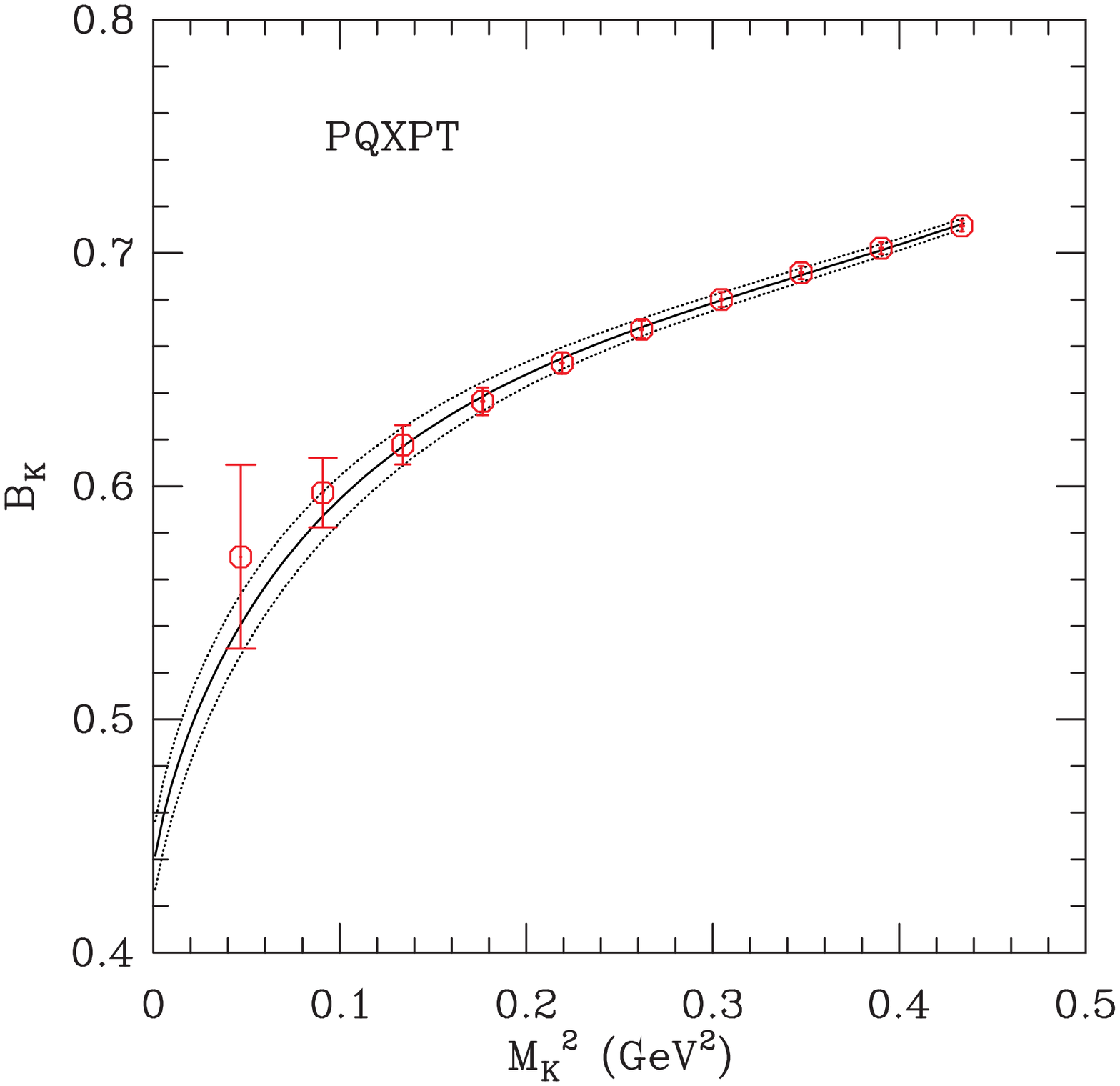}
    \includegraphics[width=.45\textwidth]
                    {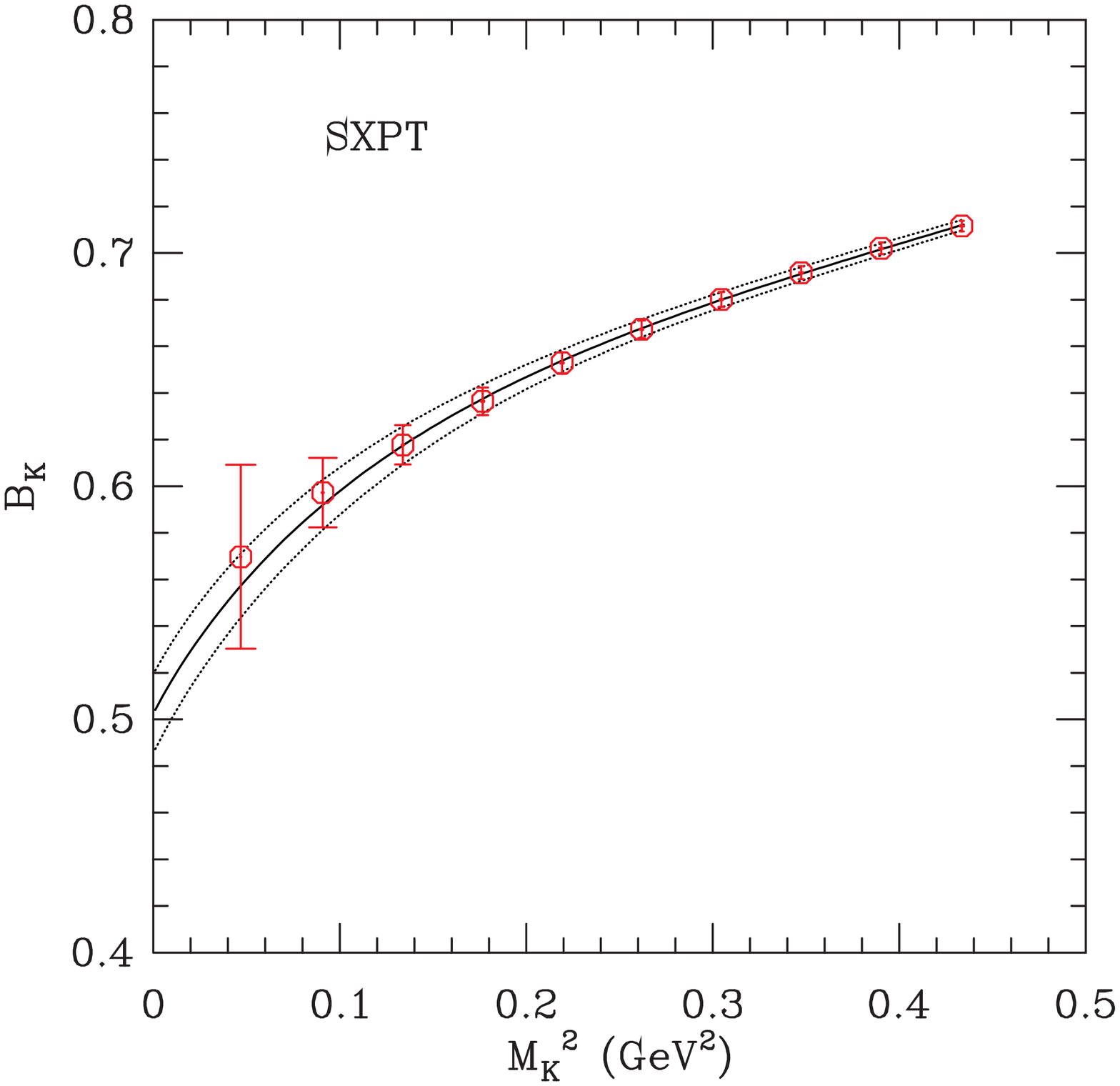}
\caption{
Fits of $B_K$ (with degenerate quarks) vs. $m_K^2$ on the MILC coarse lattices
with $am_\ell=0.01$ and $am_s=0.05$. The fits are to continuum
PQ$\chi$PT (left) and S$\chi$PT (right), as described in the text.
}
\label{fig:sxpt:1}
\end{figure}

We first try fitting to the degenerate data
without the lattice terms, 
so that only $c_{1-3}$ are non-zero.
We compare using only the Goldstone-kaon mass in loops
(``PQ$\chi$PT'' fit) to using the appropriate combination of all tastes
(``S$\chi$PT''). 
The resulting fits are shown in Fig.~\ref{fig:sxpt:1}.\footnote{%
We fit using an uncorrelated $\chi^2$, with errors in fit
parameters determined by jackknife.
Thus only the relative goodness
of fit can be estimated, but not the absolute goodness of fit.}
The S$\chi$PT fit is significantly better, 
because its smaller curvature at small quark masses more accurately
represents our data. This is the familiar
``softening'' of the chiral logarithms caused by the heavier masses of
non-Goldstone taste pions.

\begin{table}[h]
\centering
\begin{tabular}{|l||c|c|c|c|c|c|c|c|}
\hline
fit type & $c_1$ & $c_2$ & $c_3$ & $c_4$ & $c_5$ & $c_6$ & $c_7$ & $\chi^2/dof$ 
\\ \hline \hline
D-T4 & 0.39(1) & $-$.01(1) & 0.89(17) &  &  &  &  & .05(11)
\\ \hline 
D-BT4 & 0.390(3) & $-$.011(7) & 0.90(4) & .0001(4) &  &  &  & .07(16)
\\ \hline 
ND-BT4 & 0.390(2) & $-$.011(1) & 0.90(2) & .0001(2) & .13(7) 
       & $-$.011(7) & .0018(8) & .06(7)
\\ \hline 
ND-T2 & 0.31(7) & .6(5) & 0.23(50) & .002(2) & .04(4) 
      & $-$.005(4) & .0010(5) & .03(2)
\\ \hline 
\end{tabular}
\caption{Fitting parameters. Fits are described in the text.}
\label{tab:fit:1}
\end{table}
The degenerate data itself shows no indication of a logarithmically
divergent contribution (as would be produced by $f_4$). 
In a first attempt to quantify this, we try and make use of the fact
that the dominant contribution from $f_4$ is for the lightest
few kaon masses. Thus we drop the lightest mass point from the
S$\chi$PT degenerate fit, giving a fit we name ``D-T4'', 
whose parameters we list in Table~\ref{tab:fit:1}.
They have the expected magnitudes (except perhaps for $c_2$, but this
is scale dependent, and becomes $\approx 0.4$ if $\mu=0.77\,$GeV
instead of $\mu=1\,$GeV).
We then do a fit to all 10 degenerate points including $f_4$, but
with $c_{1-3}$ constrained, using the values and errors from D-T4
as Bayesian priors~\cite{ref:Bayesian}. 
This fit (``D-BT4'' in Table~\ref{tab:fit:1}), has
a very small $c_4$, with the $f_4$ term making no more than a 1\%
contribution. 
Finally, we fit to the full data set (55 points)
now including $f_{5-7}$, but with $c_{1-4}$ constrained using
the results from fit D-BT4. The resulting fit we call ``ND-BT4''.
We have also done an unconstrained fit using all 7 $f_i$ to
all 55 data points---fit ``ND-T2'' in the Table.
%
%
%

%
%

\begin{figure}[t!]
  \centering
    \includegraphics[width=.45\textwidth]
                    {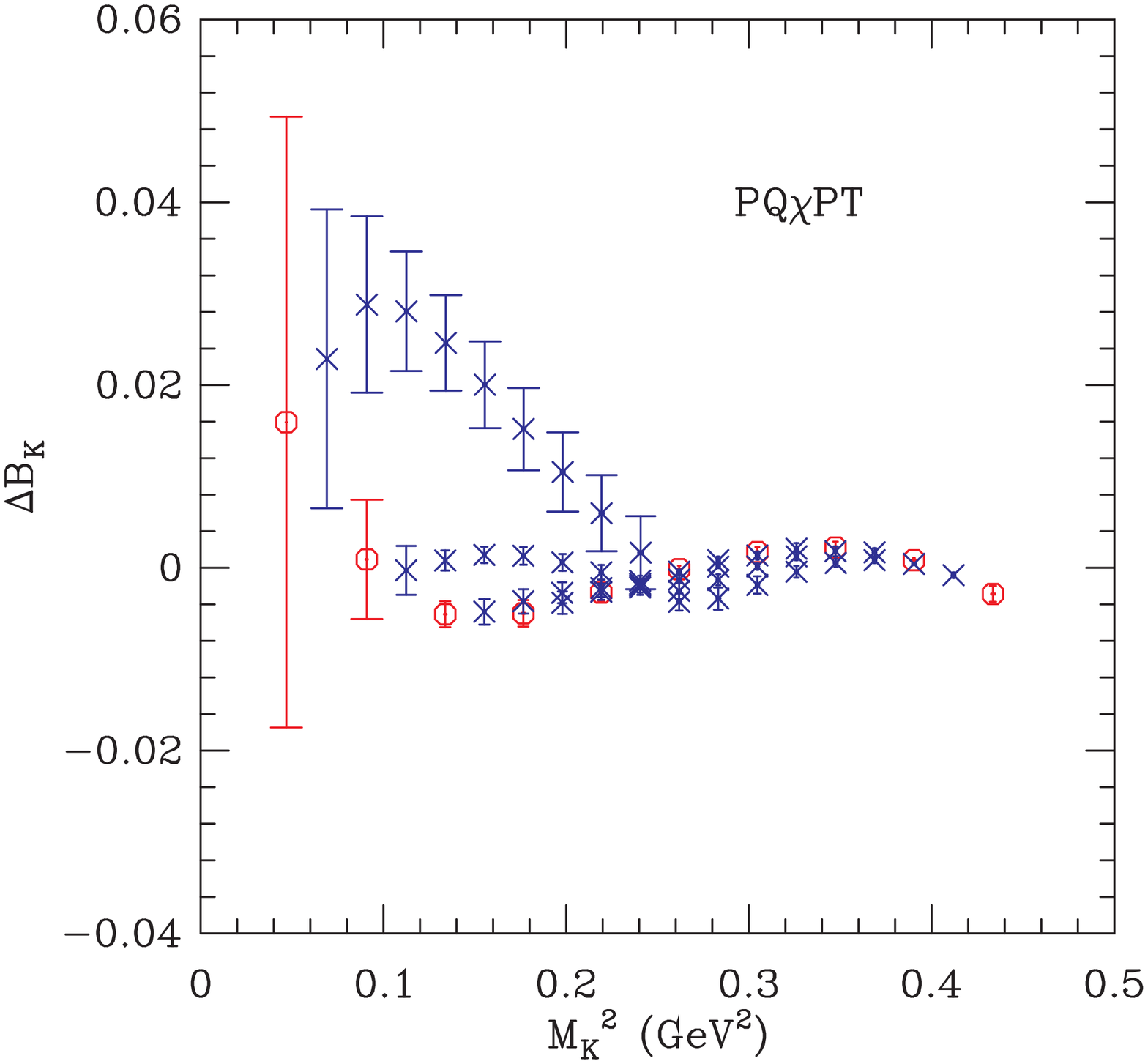}
%
    \includegraphics[width=.45\textwidth]
                    {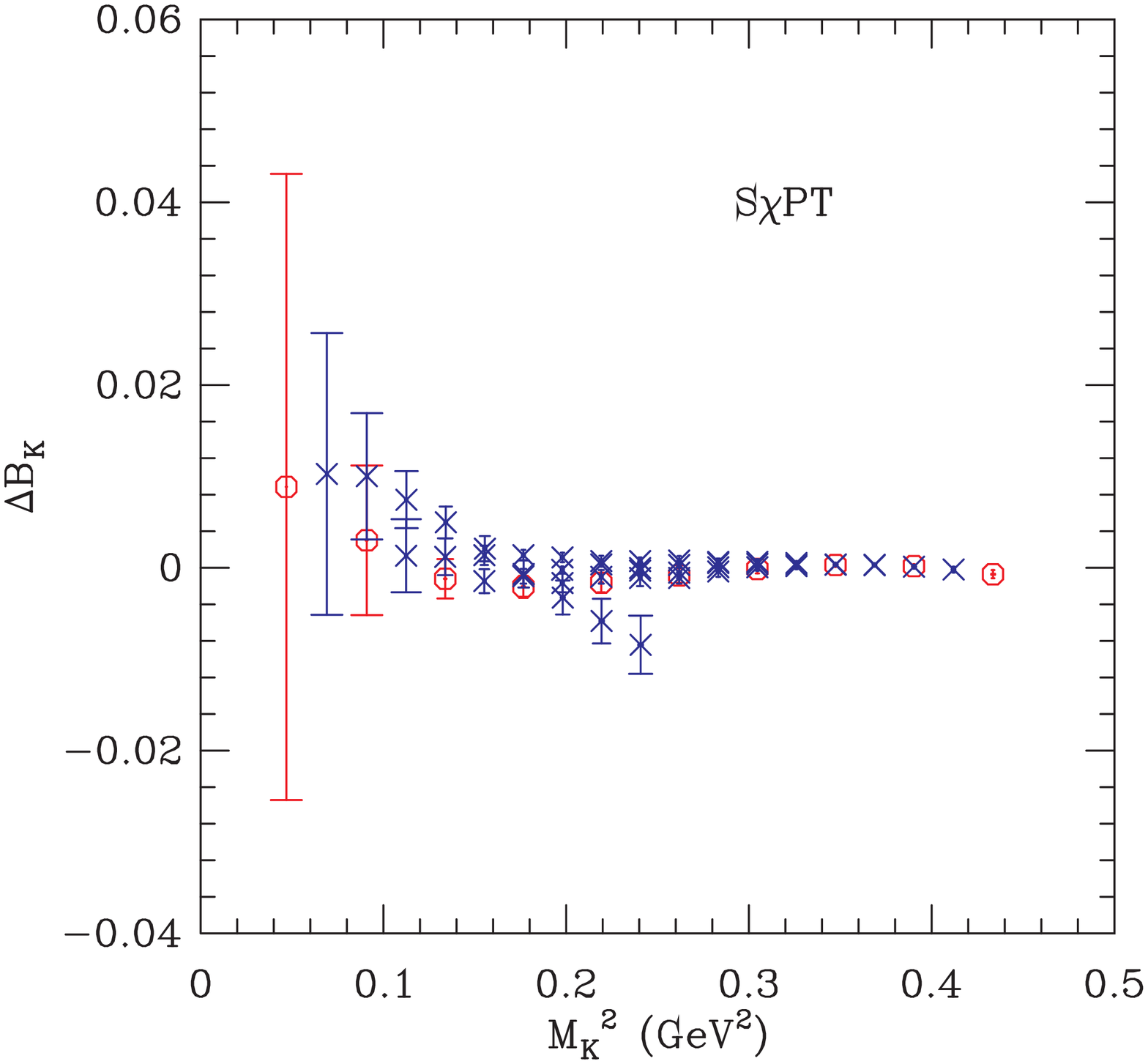}
%
\caption{$\Delta B_K$ vs. $m_K^2$, including degenerate (red) and
non-degenerate (blue) data.}
\label{fig:sxpt:4}
\end{figure}

%
The difference between 
fits ND-BT4 and ND-T2 indicates the size of the present uncertainty
in the coefficients. While this is substantial, it is encouraging
that the coefficients have sizes roughly consistent with 
the estimates (\ref{eq:guess:1}).\footnote{%
The relatively large size of $c_7$ in fit ND-BT4 is possible,
in part, because of a cancellation with the $c_6$ contribution, 
and it may be better to directly constrain both these 
coefficients to have smaller magnitudes.}
Furthermore, when we use the fit form to determine our best
estimate for the
continuum $B_K$ for physical kaon masses, we obtain consistent values:
$0.76(6)$ and $0.67(4)$.\footnote{%
These values are obtained by setting $c_{4,6,7}=0$. We stress, however,
that they still contain those discretization errors that are absorbed
into $c_1$, since they are based on fits at a single lattice spacing.}
It is also encouraging that the overall fits looks reasonable.
We illustrate this in Fig.~\ref{fig:sxpt:4}, where we compare
the residuals,
$\Delta B_K(x) = B_K (x) - f(x)$
for a continuum PQ$\chi$PT fit ($c_{4,6,7}=0$, Goldstone kaon masses only)
to the S$\chi$PT fit (ND-BT4).
We notice a significant improvement in the fitting quality using
S$\chi$PT.

We have repeated this analysis on four other coarse MILC ensembles
and one fine ensemble. Some results are collected in 
Table~\ref{tab:bk:10}. All we can conclude at this stage is that
there is a rough consistency between different coarse ensembles,
and that discretization errors are not enormous.
We stress again that these results use tree-level matching and so
are very preliminary.
\begin{table}
\begin{center}
\begin{tabular}{c | c | c | c | c | c}
\hline
$a$ (fm) & $am_l/am_s$ & geometry & ens & $B_K$( tree, ND-BT4) & $B_K$( tree, ND-T2)\\
\hline
0.12 & 0.03/0.05  & $20^3 \times 64$ & 564 &  0.83(6) & 0.63(4) \\
0.12 & 0.02/0.05  & $20^3 \times 64$ & 486 &  0.72(6) & 0.71(4) \\
0.12 & 0.01/0.05  & $20^3 \times 64$ & 671 &  0.76(7) & 0.67(4) \\
0.12 & 0.007/0.05 & $20^3 \times 64$ & 651 &  0.87(5) & 0.62(3) \\
0.12 & 0.005/0.05 & $24^3 \times 64$ & 509 &  0.80(4) & 0.66(3) \\
\hline
0.09 & 0.0062/0.031 & $28^3 \times 96$ & 995 & 0.72(4) & 0.62(3) \\
\hline
\end{tabular}
\end{center}
\caption{Comparison of tree-level $B_K$ from different ensembles (preliminary)}
\label{tab:bk:10}
\end{table}

\section{Conclusion}

We have taken the first stab at fitting our mixed-action results for
$B_K$.  We clearly have a lot of work to do to control the systematic
errors (and at this stage cannot quote a continuum result with all
errors estimated).  It is important to keep in mind, however, that the
main goal of the fitting is to provide a reasonable extrapolation
formula to the physical quark masses. We are less interested in the
coefficients themselves (except for $c_0$, which can be compared to results
from large $N_c$
approaches). We are also investigating fits based on (mixed-action
staggered) $SU(2)$ chiral perturbation theory, which appear to be much
simplified because discretization terms are of NNLO.
We also expect that the more extensive data on the fine lattices which
we are presently collecting should be more straightforward to fit.
Finally, we are improving our statistics on several of the coarse and
fine ensembles.

%
%

\section{Acknowledgments}
C.~Jung is supported by the US DOE under contract DE-AC02-98CH10886.
The research of W.~Lee is supported in part by the KICOS grant
K20711000014-07A0100-01410, by the KRF grants (KRF-2007-313-C00146 and
KRF-2008-314-C00062), by the BK21 program, and by the US DOE SciDAC-2
program.
The work of S.~Sharpe is supported in part by the US DOE grant
no. DE-FG02-96ER40956, and by the US DOE SciDAC-2 program.


\begin{thebibliography}{99}
%
\bibitem{Lellouchlat08} L.~Lellouch,
PoS (LATTICE 2008) 015.
%
\bibitem{DWFBK} C.~Allton, {\em et al.},
[arXiv:0804.0473]. 
%
\bibitem{ref:sharpe:1}  Ruth S.~Van de Water, Stephen R.~Sharpe,
Phys.~Rev.~D{\bf 73} (2006) 014003,
[{hep-lat/0507012}].
%
\bibitem{ref:wlee:11} Taegil Bae, {\em et al.},
  Phys.~Rev.~D{\bf 77} (2008) 094508, [{arXiv:0801.3000}].
%
\bibitem{Baelat08} Taegil Bae, {\em et al.}, 
  PoS (LATTICE 2008) 104, [{arXiv:0809.1219}].
%
\bibitem{msBKpaper} Weonjong Lee, {\em et al.},
  Phys. Rev. D71 (2005) 094501, [{hep-lat/0409047}].
%
\bibitem{ref:wlee:1}  Weonjong Lee and Stephen Sharpe,
Phys.~Rev.~D{\bf 60} (1999) 094503,
[{hep-lat/9905023}].
%
\bibitem{ref:bernard:1} C.~Aubin and C.~Bernard,
  Phys.~Rev.~D{\bf 68} (2003) 034014,
  [{hep-lat/0304014}].
%
\bibitem{MAChPT} O.~Baer, {\em et al.},
  Phys.Rev. D72 (2005) 054502,
  [{hep-lat/0503009}].
%
\bibitem{ref:future}  Taegil Bae, {\em et al.},
  in preparation.
%
\bibitem{ref:MILC:2004} C.~Aubin, {\em et al.}, MILC Collaboration,
  Phys.~Rev.~D{\bf 70} (2004) 114501, [{arXiv:hep-lat/0407028}].
%
\bibitem{ref:Bayesian} G.P.~Lepage, {\em et al.},
  Nucl.~Phys.~(Proc. Suppl.) B106 (2002) 12, [{arXiv:hep-lat/0110175}]
%
\end{thebibliography}
\end{document}